\documentclass[a4paper,reqno,12pt,draft]{article}
\usepackage{amssymb,euscript,bbold}
\newcommand{\be}{\begin{equation}}
\newcommand{\ee}{\end{equation}}
\newcommand{\ba}{\begin{eqnarray}}
\newcommand{\ea}{\end{eqnarray}}

\newcommand{\ft}{\footnote}

\newcommand{\nn}{\nonumber}

\begin{document}
\input{epsf}

\begin{flushright}
IC/2005/122\\
SISSA-96/2005/EP

\end{flushright}
\begin{flushright}
\end{flushright}
\begin{center}
\Large{\sc Suppressing Proton Decay in Theories with Localised Fermions}\\
\vskip 2cm

{\sc B.S. Acharya$^a$ 
 and
R. Valandro$^b$ 
}
\\
\vskip 8mm \normalsize
{\sf $^a$ Abdus Salam ICTP,
Strada Costiera 11,\\ 34014 Trieste, ITALIA\\
\smallskip
{\it bacharya@ictp.it}\\
\vskip 5mm
$^b$ International School for Advanced Studies (SISSA/ISAS),\\Via Beirut 2-4, 34014 Trieste, ITALIA \\
\smallskip
{\it valandro@sissa.it}}

\end{center}
\bigskip
\begin{center}
{\bf {\sc Abstract}}
\end{center}

We calculate the contribution to the proton decay 
amplitude from
Kaluza-Klein lepto-quarks in theories with extra dimensions, 
localised fermions and gauge fields which propagate in the bulk.
Such models naturally occur within the context of $M$ theory.
In $SU(5)$ models we show that carefully including all such modes gives a
distinctive pattern of decays through various channels including
a strong suppression of decays into neutrinos or right handed
positrons. By contrast there is no such suppression
for $SO(10)$. 

\newpage


\section{Introduction}

One of the main predictions of grand unified theories is the decay of the proton and the
experimental limits on the proton lifetime in various decay channels can
give strong constraints on GUT models. In this paper we will
study proton decay in theories with extra dimensions. In particular we will discuss theories
in which there are significantly different predictions for the proton lifetime
relative to four dimensional GUT's.
The theories of interest here are those
in which the  GUT gauge fields propagate {\it in more than} four dimensions, but the chiral matter fields are
{\it localized} in the extra dimensions.
In these cases, the GUT gauge group can be broken to that of
the Standard Model through compactification; for example by a gauge field expectation value in the extra
dimensions.
We will show that in such models one can get an enhancement of the
lifetime in some decay channels with respect to the four dimensional
GUT prediction.

The mechanism for this is the following. Firstly the symmetries of the model are such that
dimension five baryon number violating operators are suppressed. This is natural for instance in certain
$M$ theory vacua of this kind \cite{decon}. The leading contribution at dimension six is through the
mediation of colour triplet heavy gauge bosons. There is an infinite, Kaluza-Klein tower of such massive lepto-quarks\ft{These are analagous to the $X$ and $Y$ bosons of four dimensional
$GUT$'s the difference being in the number of such particles.}.
Then, since generically (in the language of $SU(5)$)
the points where matter ${\bf 10}$'s are localised are distinct from the points supporting ${\bf \bar{5}}$'s,
there is a qualitative difference between the decay modes such as $p\rightarrow \pi^0e^+_L$ and those such as
$p\rightarrow \pi^0e^+_R$ or $p\rightarrow \pi^+ \bar{\nu}_R$. The reason is simple: the first decay mode
comes from a current-current correlator where  {\it both} fermion currents are of a single ${\bf 10}$
multiplet localised at the {\it same} point in the extra dimensions; on the other hand for the
other two channels the two currents involve a ${\bf {\bar 5}}$ and
${\bf 10}$ multiplet which are localised at {\it different} points in the extra dimensions.
The propagator for the Kaluza-Klein lepto-quarks in the extra dimensions can, as we will see
explicitly, take a non-trivial form. The fact that the value of the propagator can become small, even zero, is what suppresses the latter
two decay channels. In other words, cancellations to the amplitudes occur by including the contribution
of all the relevant Kaluza-Klein modes.

For the first channel (as we will review), the two currents are at the same point and the universal short distance behaviour of the propagator
leads to a divergence in the amplitude, which in the $M$ theory context studied in \cite{wittfried}
was argued to be regularised. For the second and third channels studied here, this divergence is absent in $SU(5)$
precisely because the two currents involved are separated in the extra dimensions.
Hence there can be a suppression of $p\rightarrow \pi^0 e^+_R$ and
$p\rightarrow {\bar \nu}_R$.
For the case of $SO(10)$ where all the matter of one generation resides in a single ${\bf 16}$ multiplet,
all three channels suffer
the {\it same} divergence, hence we do not expect any qualitative difference between the three amplitudes
in $SO(10)$.
This gives a simple way to distinguish $SO(10)$ from $SU(5)$. Similarly decays involving
more than one generation e.g. $p \rightarrow K^0 \mu^+$ can also be suppressed by the
small value of the propagator; this suppression can also occur in $SO(10)$ because two
different ${\bf 16}$ currents are involved.

Of course,  the detailed prediction for the cross-section for the proton decay involving
currents in different multiplets is quite model dependent, since it depends both on the
particular metric on the extra dimensions and on the precise locations of the two currents involved in the decay.
To investigate this model dependence we calculated  the amplitude in a variety of different spaces.
In particular we took $Q$ to be a space with constant positive, zero or negative curvature and showed
that a significant effect can always occur.

\smallskip

In the next section we review the basic calculation of the current-current correlator in theories
with localised fermions following \cite{wittfried}.
We then go on to calculate the Green's functions for some model three dimensional spaces
and show explicitly that the propagator can
become very small depending on where the fermions are localised.
The final section contains our conclusions and a comparison
between results obtained for the $M$ theory models and some other
extra dimensional GUT's such as \cite{hebmarch}.

\newpage


\section{Proton Decay in Extra Dimensions with Localized Fermions.}

We will consider theories in which the fermions and Higgs particles of the Standard Model are localised
in the extra dimensions, but in which the gauge fields propagate in (part of) the bulk. The full spacetime
is thus of the form $X \times M^{3,1}$ with $M^{3,1}$ our four dimensional spacetime and $X$ the
compact extra dimensions. The Standard Model matter particles are localised at points on $X$ and
the gauge fields propagate along a submanifold $Q$ of $X$ times the four dimensional spacetime.
In the GUT context, the GUT gauge group could be broken to $SU(3) \times SU(2) \times U(1)$ by a
Wilson loop of the gauge field on $Q$. We will also restrict our attention to theories in which
the leading contribution to the violation of baryon number comes from dimension six operators (the analog
of the gauge boson contribution in the original non-supersymmetric four-dimensional GUT's).

Although our results are more generally applicable, we will for concreteness focus on the case of $M$ theory
compactifications on manifolds of $G_2$-holonomy which provide an explicit realisation of theories of this kind.
Here $X$ is a 7-manifold with $G_2$-holonomy, $Q$ is a three
dimensional submanifold along which $X$ has a particular orbifold singularity \cite{bsa}, and the chiral fermions are
localised at particular kinds of conical singularity \cite{Acharya:2001gy}\ft{These matters
are reviewed in \cite{physrep}.}. The fact that dimension four and five baryon number violating operators
are naturally supressed in such models was explained in \cite{decon}.

Also, for definiteness we will restrict attention to the case where the GUT gauge group is $SU(5)$, so that
before turning on the Wilson loop which breaks the symmetry to the Standard Model gauge group, each generation
of (supersymmetric)
Standard Model matter resides in the ${\bf \bar{5} \oplus 10}$ with Higgs particles in the ${\bf {\bar 5}  \oplus 5}$.
So with the minimal field content there are eight points $P_i \subset Q$ where matter is localised:
two for the Higgs multiplets, three for
the ${\bf 10}$ matter and three for the anti-fundamental generations.

Following \cite{wittfried} we now describe how the Greens function on $Q$ appears in the calculation
of the proton decay amplitude at dimension six in theories of this kind.

A matter current which can absorb or emit a massive gauge boson is of the form
\be
J_{\mu} = J_{\mu}^{\bf \bar{5}} + J_{\mu}^{\bf 10}
\ee
where the subscripts indicate the origin of the particles involved.

In the case of four dimensional GUT theory, the gauge boson contribution to the proton
decay amplitude is essentially
\be
g_{GUT}^2 \int d^4 x J_{\mu}(x) \tilde{J}^{\mu}(0) D(x,0)
\ee
where $J$ and $\tilde{J}$ are the two currents involved and $D$ is the propagator of the massive gauge
boson. The latter transforms as $(\bf{3},\bf{2})^{-5/3}$ under the Standard Model gauge symmetry. Since
the size of the proton is much bigger than the integration region which gives the dominant contribution, we can replace
$J^\mu(x)$ by $J^\mu(0)$. In this case the integral gives the result
\be\label{ampl4}
{g_{GUT}^2 J_{\mu}(0) \tilde{J}^{\mu}(0) \over M^2}
\ee
with $M$ the boson mass. This is a consequence of the equation for the propagator
\be\label{prop4}
(\Delta_{4} + M^2 )D(x,0) = \delta(x,0)
\ee
where $\Delta_{4}$ is the  four dimensional Laplacian.
In the higher dimensional theories under discussion here one must also include
the contribution of all charged Kaluza-Klein modes in the $(\bf{3},\bf{2})^{-5/3}$ representation.
In these cases the propagator $D(x, y;x', y')$ is a function of the coordinates $y$ on $Q$ as well
as those $x$ on $M^{3,1}$ and the currents are functions of $x$ but are labelled by the points $P_i$
which are the values of $y$ where the matter particles are located. So we get an amplitude of the form
\be
g_7^2 \int d^4 x J_{\mu}(x, {P_1}) \tilde{J}^{\mu}(0, {P_2}) D(x,P_1 ; 0, P_2 )
\ee
Again we can replace $J^\mu(x)$ by $J^\mu(0)$, so the previous expression is well approximated by
\be\label{ampl7}
g_7^2  J_{\mu}(0, {P_1}) \tilde{J}^{\mu}(0, {P_2})\int d^4 x D(x,P_1 ; 0, P_2 )
\ee

The difference between (\ref{ampl4}) and (\ref{ampl7}) is
the $P_i$ dependent function:
\be\label{GG}
   G(y_1,y_2)\equiv \int_{M^{3,1}} d^4x D(x,y_1;0,y_2).
\ee

The seven dimensional propagator satisfies
\be\label{prop7}
(\Delta_{4} + \Delta_{Q})D(x,y_1;0, y_2) = \delta(x,0)\delta (y_1 , y_2)
\ee
where $\Delta_{Q}$ is the gauge covariant Laplacian on $Q$. From this we see that the eigenvalues of $\Delta_Q$ act as masses$^2$ from the four dimensional viewpoint.

$D(x,y_1;0,y_2)$ is the contraction of the Feynman propagator on $M^{3,1}\times Q$ of the seven dimensional
gauge fields in the $(\bf{3},\bf{2})^{-5/3}+(\bf{\bar{3}},\bf{2})^{+5/3}$ representation of the Standard Model gauge group:
\be
   D(x,y_1;0,y_2)=\frac{1}{(2\pi)^4}\sum_k\int d^4p \frac{e^{-i p\cdot x} \bar{\Psi}_k(y_1)\Psi_k(y_2)}{-p^2+\lambda_k}
\ee
where $\Psi_k$ are the eigenfunctions on $Q$ of $\Delta_Q$
with eigenvalues $\lambda_k\leq 0$, and the integral over 
$p$ is considered after euclidean continuation.

When there are no zero modes of the Laplacian on $Q$, one can substitute this expression in (\ref{GG}) and get:
\be\label{GG1}
   G(y_1,y_2)=\sum_k \frac{\bar{\Psi}_k(y_1)\Psi_k(y_2)}{\lambda_k}
\ee
{\it ie} the Green's function of the scalar Laplacian on $Q$ for scalar fields valued in  $(\bf{3},\bf{2})^{-5/3}$ 
representation.

When there is a non zero background gauge field such that  the $SU(5)$ symmetry is broken to the Standard Model 
gauge group, the Laplacian typically
has no zero modes in the space of functions with values in $(\bf{3},\bf{2})^{-5/3}+(\bf{\bar{3}},\bf{2})^{+5/3}$ and the expression
(\ref{GG1}) is well defined.

From the decomposition of the operator product
\be
J_{\mu}\tilde{J}^{\mu} = J^{\bf 10}_{\mu} \tilde{J}^{\mu{\bf 10}} + J^{\bf 10}_{\mu} \tilde{J}^{\mu{\bf \bar{5}}}
+ J^{\bf \bar{5}}_{\mu}\tilde{J}^{\mu{\bf 10}} + J^{\bf \bar{5}}_{\mu}\tilde{J}^{\mu{\bf \bar{5}}}
\ee
only the first term contributes to the cross-section for the decay of the proton into
left-handed positrons.
The second and third contribute to the decays into neutrinos whereas the last term does not contribute
to the decay.  So for the decays modes such as $p\rightarrow \pi^0e^+_L$  studied in \cite{wittfried}  both
${\bf 10}$ currents are localised at the same point on $Q$. 
The corresponding Greens function in (\ref{GG1})
is therefore evaluated at $P_1 = P_2$ for this decay channel and 
therefore the classical formula is {\it divergent}\footnote{
Note that in the cases when $Q$ is one dimensional, the Green's function is 
not divergent when $P = P'$. This actually happens in some orbifold 
GUT models \cite{hebmarch}}. 
 This is presumably
regularised in $M$ theory \cite{wittfried}.

However, since generically the points supporting the ${\bf \bar{5}}$ and the ${\bf 10}$ are distinct (for example to generate
reasonably small Yukawa couplings), for the decay channels involving neutrinos, the ${\bf 10}$ is at a point $P_1$ distinct from the point $P_2$ supporting the ${\bf \bar{5}}$ current.
Therefore the current-current correlator depends explicitly on the Green's function on $Q$ evaluated
at two different points $G(P_1 ; P_2)$. When $G(P_1 ; P_2)$ takes a small value
the decay of the proton into neutrinos is suppressed accordingly. Generically, $Q$ is
a curved, compact manifold and the Green's function will be a non-trivial function of the
geodesic distance $d(P_1 , P_2)$ between the points. In order to investigate the behaviour
of such functions, in particular, whether or not they can take small values, we will present
some explicit sample calculations
in the $M$ theory context.

\newpage


\subsection{M Theory compactifications on G2 manifolds.}

In the $M$ theory context, the gauge fields propagate on a three-dimensional
subspace $(Q)$ of the bulk. 
If $Q$ has incontractible loops, so that its fundamental group $\pi_1(Q)$ is non-empty, 
it is possible to break $SU(5)$ to the 
Standard Model gauge group by a Wilson line in the vacuum. 
This modifies the Kaluza-Klein spectrum with respect to zero background gauge field; 
for example the lightest modes of the gauge fields 
corresponding to the unbroken generators remains massless,  while the others generically
get a non-zero mass.

For an example, we take $Q=\mathcal{S}^3/\Bbb{Z}_p$ \cite{wittfried}. This space has non-contractible circles which
correspond to open curves in $\mathcal{S}^3$ 
that connect two points identified by the elements of $\Bbb{Z}_p$. The background gauge field can be taken
to be a Wilson line around such circles. For instance, the following Wilson line breaks $SU(5)$ to the Standard
Model gauge group:

\be\label{UG} 
   U_\Gamma=P \,e^{i\oint_{\gamma_\Gamma} A_{\rm bkg}}=\left(
	\begin{array}{ccccc}
		e^{4\pi i q/p}&&&&\\& e^{4\pi i q/p}&&&\\&&e^{4\pi i q/p}&&\\&&&e^{-6\pi i q/p}&\\&&&&e^{-6\pi i q/p}\\
	\end{array}\right).
\ee

If $\Phi(y)$ is a scalar charged under the gauge symmetry then $\Phi(y) = \Phi(y + 2\pi R)$ where
we took $y$ to be the coordinate around the loop. The Laplacian acting on $\Phi(y)$ depends explicitly
on the background gauge field. This makes computing the spectrum difficult. However, since the background
gauge field has zero field strength, $F=0$, we can locally
eliminate the gauge
field dependence by performing a non-single valued gauge transformation $g(y)$ (see \cite{Hall:2001tn} for
a simple example). The price we pay for this is 
to change the periodicity condition on $\Phi(y)$ to
\ba\label{ModBoundCond}
   \Phi(y)=U_\Gamma \Phi(\Gamma y) & {\rm where} & \Gamma\in\Bbb{Z}_p \mbox{   and   } y\in \mathcal{S}^3 .
\ea
where $U_\Gamma = g(2\pi R)$ acts in the appropriate representation. 

Thus, in the presence of the Wilson line, a charged scalar field on $Q=\mathcal{S}^3/\Bbb{Z}_p$ is equivalent to a field on
$\mathcal{S}^3$ satisfying the above invariance conditions. Since the spectrum of the ordinary Laplacian is known
on the round $\mathcal{S}^3$ we can proceed.

In order to compute the Green's function $G(y_1,y_2)$, one needs the eigenmodes of the Laplacian on $Q$
which satisfy the boundary conditions (\ref{ModBoundCond}) and 
which take values in the adjoint representation of $SU(5)$. The decomposition of this representation under the 
group $SU(3)_c\times SU(2)_L\times U(1)_Y$ is given by:
\be\label{decomp}
\bf{24}=(\bf{8},\bf{1})^0+(\bf{1},\bf{3})^0+(\bf{1},\bf{1})^0+(\bf{3},\bf{2})^{-5/3}+(\bf{\bar{3}},\bf{2})^{+5/3}
\ee

We will take $U_\Gamma$ such that the vector spaces of the decomposition (\ref{decomp}) 
are eigenspaces of this transformation.
In particular, the Wilson line is chosen to leave
the $(\bf{8},\bf{1})^0+(\bf{1},\bf{3})^0+(\bf{1},\bf{1})^0$ part to be invariant, since the Standard Model
gauge symmetry is unbroken. In particular we are interested in the scalar fields in the representation
$(\bf{3},\bf{2})^{-5/3}+(\bf{\bar{3}},\bf{2})^{+5/3}$ which obey (\ref{ModBoundCond}).

\smallskip

We will now compute the Green's function explicitly in several examples when $Q$ has constant curvature.
The details of most of these computations are given in the Appendix, but we will give some explicit
derivations below also.


\subsection*{Constant Positive Curvature}

3-manifolds with constant positive curvature are all quotients of the round 3-sphere by a discrete group.
We will compute the relevant Green's function for quotients by $\Bbb{Z}_p$, beginning with the simplest
example.


\subsubsection*{The simplest case: $\Bbb{RP}^3=\mathcal{S}^3/\Bbb{Z}_2$}

This is a particular case of the example presented above, in which $p=2$, $q=1$ and
\be\label{UG1} 
U_\Gamma=\left(
\begin{array}{ccccc}
  1&&&&\\& 1&&&\\&&1&&\\&&&-1&\\&&&&-1\\
\end{array}\right).
\ee
Under this transformation the generators of the $(\bf{3},\bf{2})^{-5/3}+(\bf{\bar{3}},\bf{2})^{+5/3}$ representation
are odd (because the adjoint of the Standard Model is the only invariant representation). Therefore
to get invariant eigenmodes on $\mathcal{S}^3/\Bbb{Z}_2$ we have to take the odd eigenfunctions on
$\mathcal{S}^3$
under the $\Bbb{Z}_2$ transformation.

The eigenvalues of the Laplacian on $\mathcal{S}^3$ are labelled by integers $k$ and
given by $\lambda_k=-k(k+2)$. The relative eigenspaces are
\be
\mathcal{V}^k=\{T_{k; m_1,m_2} \,\, |-k/2\leq m_1,m_2\leq k/2\}
\ee
where
\be
T_{k; m_1,m_2}(\chi,\theta,\varphi)=\sqrt{\frac{k+1}{2\pi^2}}\,\, D^{k/2}_{m_2,m_1}(\chi,\theta,\varphi)
\ee
where $D^{k/2}_{m_2,m_1}$ are the Wigner $D$-functions, written in terms of angular coordinates on $SU(2)$.
The $D$'s are just the matrix elements of the spin
$k/2$ representation of $SU(2)$.

Under a $\Bbb{Z}_2$ transformation, $T_{k; m_1,m_2}(y)\mapsto (-1)^k T_{k; m_1,m_2}(y)$. So the odd eigenfunctions are those 
relative to odd $k$. We have also to change the normalization of such functions, because the volume of $\mathcal{S}^3/\Bbb{Z}_2$ 
is half of the volume of the defining $\mathcal{S}^3$.

The sum (\ref{GG1}) becomes:
\be\label{GGRP}
G(y_1,y_2)=\frac{1}{\pi^2}\sum_{k=1,3,...}^\infty\frac{k+1}{-k(k+2)}
\sum_{m_1,m_2}\bar{D}^{k/2}_{m_1,m_2}(g(y_1)) D^{k/2}_{m_1,m_2}(g(y_2))
\ee
From group theory we know that \cite{fink}:
\be\label{DDgr}
\sum_{m_1,m_2}\bar{D}^{k/2}_{m_1,m_2}(g(y_1)) D^{k/2}_{m_1,m_2}(g(y_2))  = \frac{\sin[(k+1) d(y_1,y_2)]}{\sin[d(y_1,y_2)]}
\ee
where $d(y_1,y_2)$ is the geodesic distance on the 3-sphere between $y_1$ and $y_2$.

Inserting this relation in (\ref{GGRP}) one can do the sum explicitly:
\ba\label{GGRP1}
G(y_1,y_2)&=&\frac{1}{\pi^2}\sum_{k=1,3,...}^\infty\frac{k+1}{-k(k+2)} \frac{\sin[(k+1)d ]}{\sin[d ]} \nn\\
			&=& \frac{1}{\pi^2}\sum_{j=0}^\infty\frac{2j+2}{-(2j+1)(2j+3)} \frac{\sin[(2j+2)d ]}{\sin[d ]}\nn\\
			&=& -\frac{1}{2\pi^2 \sin d}\left(\sum_{h=1}^\infty\frac{h}{h^2-1/4} \sin[2hd ]\right)\nn\\
			&=& -\frac{1}{2\pi^2 \sin d}\left( \frac{\pi}{2}\frac{\sin(\pi/2-d )}{\sin (\pi/2)}\right)\nn
\ea
where we used \cite{integrals}
and doing the last step, one gets:
\be\label{GGRP3}
G(y_1,y_2) = -\frac{1}{4\pi}\frac{1}{\tan d(y_1,y_2)}
\ee
where $d(y_1,y_2)\in[0,\pi/2]$ is restricted to the points representing
$\mathcal{S}^3/\Bbb{Z}_2$. We see that the absolute value of the Green's function takes all  values between $0$ 
and $\infty$. So in this example, if the ${\bf 10}$ multiplet and the ${\bf \bar{5}}$ multiplet are maximally
separated in $\Bbb{RP}^3$ the Green's function is zero and the cross-section vanishes. In this case the lifetime of
the decay channel into neutrinos receives no contribution at all from dimension six operators.


\subsubsection*{General Lens Space}

The Lens space  $L(p,r)$ is the quotient of the
3-sphere by the cyclic group whose generator 
$\Gamma$ is the $SO(4)$ isometry given in $\Bbb{R}^4$ by \cite{Weeks}:
\be\label{Gamma} 
\Gamma=\left(
\begin{array}{cccc}
  \cos(2\pi/p)&-\sin(2\pi/p)&&\\ \sin(2\pi/p) &\cos(2\pi/p)&&\\
  &&\cos(2\pi r/p)&-\sin(2\pi r/p)\\&&\sin(2\pi r/p)&\cos(2\pi r/p)\\
\end{array}\right)
\ee
$U_\Gamma$ is given by (\ref{UG}). With the same procedure used for the previous case, one obtains 
the formula for the Green's function:
\be\label{GGlens}
G(y_1,y_2) = \sum_{w=1}^p  u^w \frac{d(y_1,\Gamma^w y_2)-\pi}{4\pi^2 \tan d(y_1,\Gamma^w y_2)}
\ee
where $u\equiv e^{2\pi i 5wq/p}$, and $d\in[0,\pi]$ is again the geodesic distance on the sphere.

In order to study (\ref{GGlens}), we use the cartesian coordinates on $\Bbb{R}^4$ 
where $\mathcal{S}^3$ is defined by $x^2+y^2+z^2+t^2=1$, and
choose, without loss of generality, $y_2=y_O\equiv(1,0,0,0)$.

At first, we note that it has a singularity only at $y_1 \rightarrow y_O$, at which 
$d\rightarrow 0$. In this limit $G\sim \frac{1}{4 \pi d}$, as one expects. One can check that
this is the only divergence.
Secondly, we note that the Green's function on a Lens space has always zeros. Actually,  
the points  $\tilde{y}_1 = (0,0,z,t)$ (with $z^2+t^2=1$)
have the same distance $d=\pi/2$ from each of the points  $\Gamma^w y_O=(x,y,0,0)$. This is because 
the distance on the sphere is given by $\cos d=1-\frac{d_E^2}{2}$ in terms of the euclidean 
distance on $\Bbb{R}^4$, and the chosen points have always $d_E^2=2$. So. for this value of $d$
\be
G=\frac{d-\pi}{4\pi^2 \tan d}\sum_w u^w=0
\ee

\newpage 

\subsection*{Constant Zero Curvature}

Any closed, compact zero curvature manifold is a quotient of the flat 3-torus by a discrete group. Here
we consider the case of the torus itself.


\subsubsection*{The 3-dimensional torus}

We consider the square torus, with coordinates $\vec{x}$ and $-1/2\leq x_i<1/2$. 
It is a non-simply connected manifold, whose fundamental group 
has three generators. We choose a background gauge field such that the 
holonomy associated to each of three generators is
given by 
\be
U_i=\left(
\begin{array}{ccccc}
  1&&&&\\& 1&&&\\&&1&&\\&&&-1&\\&&&&-1\\
\end{array}\right).
\ee
with $i=1,2,3$.
This choice breaks $SU(5)$ to the Standard Model gauge group. \\
The eigenfunctions on the torus with values in $(\bf{3},\bf{2})^{-5/3}+(\bf{\bar{3}},\bf{2})^{+5/3}$ 
are those which satisfy the boundary conditions:
\be\label{BoundCondTor}
\Phi(\vec{x})=(-1)^{\sum_i k_i}\Phi(\vec{x}+\vec{k})
\ee
for arbitrary $\vec{k}$ with $k_i\in \Bbb{Z}$. This is because each lattice generator acts as $-1$
in the representation $(\bf{3},\bf{2})^{-5/3}+(\bf{\bar{3}},\bf{2})^{+5/3}$. 

Once we have found them, we can compute the Green's function, obtaining:
\be\label{GreenTorus}
G(\vec{x},\vec{0})=- \sum_{\vec{m}} \frac{(-1)^{\sum_i m_i}}{4\pi|\vec{x}-\vec{m}|}
\ee
This is the same formula as the electrodynamic potential of a distribution of positive and
negative charges situated on nodes of the lattice given by
    $\vec{m}$, where the sign of the charge is given by $(-1)^{\sum_i m_i}$. It has the 
expected $\frac{1}{4\pi|\vec{x}|}$ singularity when $\vec{x}\sim\vec{0}$.
Moreover it has zeros when any of the $x_i$ is equal to $1/2$. 
Actually, the charges can be grouped
in pairs, one negative, one positive each of which has
the same distance from such points.
Summing all these contributions gives so zero since the contribution from each pair
is zero. One can check this more explicitly by evaluating the expression
(\ref{GreenTorus}) in the case $\vec{x}=(1/2,x_2,x_3)$.

\newpage

\subsection*{Constant Negative Curvature}

A constant negative curvature 3-manifold is a quotient of hyperbolic 3-space $\Bbb{H}^3$
by a discrete group. In the compact case such groups are very rich and complicated and a description
of the eigenfunctions of the Laplacian on charged scalars is difficult to give explicitly.
Instead of attempting an explicit computation, 
we will compute the Green's functions on $\Bbb{H}^3$ itself and we will give an argument for 
the large suppression of the Green's funcion on compact manifolds with negative curvature.

\subsubsection*{The Hyperbolic 3-space}

In this case we get the Green's function, by computing the Heat Kernel $H(y_1,y_2;t)$ 
and then integrating on $t$. Actually
\be\label{HeatK}
H(y_1,y_2;t)=\sum_k e^{-|\lambda_k| t}\bar{\Psi}_k(y_1)\Psi_k(y_2)
\ee
and, if the integral converges,
\ba
\int_0^\infty dt H(y_1,y_2;t)&=&  \int_0^\infty dt \sum_k e^{-|\lambda_k| t}\bar{\Psi}_k(y_1)\Psi_k(y_2) \nn\\
&=& -\sum_k \frac{\bar{\Psi}_k(y_1)\Psi_k(y_2)}{\lambda_k}\nn\\
&=& -G(y_1,y_2)
\ea
Following the explicit computation reported in the Appendix, one gets:
\be
G_{\Bbb{H}^3}(y_1,y_2)=-\frac{1}{4\pi}\frac{e^{-d(y_1,y_2)}}{\sinh d(y_1,y_2)}
\ee
In this case the Green's
function is suppresed already at distance of order $1$.

The Green's function on a quotient of $\Bbb{H}^3$ by a discrete group in the presence
of Wilson loops will be an infinite sum of the type:
\be
G(y_1,y_2)=-\frac{1}{4\pi}\sum_\Gamma u(\Gamma)\frac{e^{-d(y_1,\Gamma y_2)}}{\sinh d(y_1,\Gamma y_2)}
\ee
For the Torus we have found a similar expression and we have seen that it has zeros. In this case
we have also the suppression of $G_{\Bbb{H}^3}$ at distance of order $L=V^{1/3}$, where $V$ is the volume of
the final compact manifold. 
So it is conceivable that the combined action of the
cancellation by the Wilson lines phases and the exponential suppression will bring $G(y_1,y_2)$, 
if not to have zeros, to be strongly suppressed for particular choices of the points $(y_1,y_2)$. This would allow us to 
make the same conclusions as for the previous cases.

\newpage

\section{Conclusions}

We have found that in $SU(5)$ theories,
that the decays of protons into pions and neutrinos or right handed
positrons
{\it can} be highly suppressed. So if, for instance,
protons are observed to decay into positrons and the lifetime
for the decay channel into neutrinos is established to be significantly
longer than this decay time, the mechanism described here offers a
natural explanation. Unfortunately super-Kamiokande is not sensitive
to the helicity of outgoing positrons. A measurement of the dominant
helicity would be a strong test of these  models with localised fermions
and should be considered when planning future proton decay experiments.
Furthermore the qualitative difference between
$SU(5)$ and $SO(10)$ that we noted in the introduction is quite
striking and seems to go beyond what can be explained simply in
four dimensional field theory with a finite spectrum, although perhaps
the results described here can also be ``deconstructed'' analagously to
some of the results in \cite{decon}.

In the models described here {\it all} the fermions of the standard model
are {\it localised} in the extra dimensions. In this case, the a priori problem
that the $SU(5)$ mass relations for the first two generations are incorrect
can be solved by introducing
additional vector-like localised matter (eg ${\bf 5 \oplus {\bar 5}}$) which
mix with these generations \cite{decon}. In many other models considered
in the literature, where proton decay has been considered in detail 
\cite{hebmarch},
this problem can be solved by including fermions in the bulk of $Q$
which then mix with the localised fermions.
In $M$ theory this option is not obviously 
available.
Furthermore, in the models of the sort considered in \cite{hebmarch},
the extra dimensions have boundaries and $SU(5)$ is broken by boundary conditions.
These two considerations can then lead to models in which  decay channels involving
the {\it first generation only} are absent at dimension six. The dominant decays are then those
such as $p \rightarrow K^0 \mu^+$. By contrast, in the models under consideration in
this paper decays inolving
the first generation are allowed. Moreover, as we have explained, the same mechanism
which suppresses, say, $p \rightarrow \pi^+ {\bar \nu}_R$ can also suppress
$p \rightarrow K^0 \mu^+$.
In principle therefore it is straightforward to distinguish between these
different types of models experimentally.


\vskip1cm

{\large \sf Acknowledgements}

\vskip3mm

We would like to thank L. Boubekeur, P. Creminelli, J. David, G. Milanesi, A. Smirnov
and G. Thompson for valuable discussions.

\newpage

\appendix

\section{Green's function on Lens spaces: details}

In order to compute the Green's function on Lens spaces, one needs the eigenmodes on them.

\subsection{Eigenmodes of Laplacian on the 3-sphere}

In order to study the eigenmodes of the Laplacian on Lens spaces, we need to review the eigenmodes on the 3-sphere 
\cite{Weeks}.

At first, we introduce the toroidal coordinates on the 3-sphere $\mathcal{S}^3$. 
Let $x,y,z$ and $t$ be the usual coordinates in $\Bbb{R}^4$, so $\mathcal{S}^3$ is defined by $x^2+y^2+z^2+t^2=1$, and can be 
parametrized by the coordinates $\chi,\theta$ and $\varphi$ as
\ba\label{TorCoord}
x&=&\cos\chi\,\,\cos\theta\\
y&=&\cos\chi\,\,\sin\theta\\
z&=&\sin\chi\,\,\cos\varphi\\
t&=&\sin\chi\,\,\sin\varphi
\ea
with $0\leq\chi\leq\pi/2$, $-\pi \leq \theta\leq \pi$ and $-\pi \leq\varphi\leq \pi$.

The eigenvalues of the Laplacian on $\mathcal{S}^3$ are given by $\lambda_k=-k(k+2)$. The relative eigenspaces are given by
\be
\mathcal{V}^k=\{T_{k; m_1,m_2} \,\, |-k/2\leq m_1,m_2\leq k/2\}
\ee
where the $T$'s can be expressed in terms of the Wigner $D$-functions $D^{k/2}_{m_2,m_1}$:
\be\label{TDapp}
T_{k; m_1,m_2}(\chi,\theta,\varphi)=\sqrt{\frac{k+1}{2\pi^2}}\,\, D^{k/2}_{m_2,m_1}(\chi,\theta,\varphi)
\ee

\subsection{Eigenmodes of Laplacian on Lens Spaces}

The Lens space $L(p,r)$ is the quotient of the 3-sphere by the cyclic group whose generator $\Gamma$ 
is the isometry \cite{Weeks}
\ba\label{TrZpqapp}
\chi \mapsto \chi; &   \theta \mapsto \theta + 2\pi/p; &\varphi \mapsto \varphi + 2\pi r/p
\ea
It can be described using toroidal coordinates, with limit 
$0\leq \chi\leq \pi/2$, $-\pi/p < \theta < \pi/p$ and $-\pi r/p < \varphi < \pi r/p$.
Obviously it cannot be covered only with  one such patch, but the set of non-covered points is of null mesure.
Moreover it gives a good local description around 
the point $(\chi,\theta,\varphi)=(0,0,0)$.

Under the transformation (\ref{TrZpqapp}) the eigenfunctions found above transform as:
\be\label{Ttransfapp}
T_{k; m_1,m_2}(\chi,\theta,\varphi)  \mapsto  e^{2\pi i(\ell+m r)/p} \,T_{k; m_1,m_2}(\chi,\theta,\varphi)
\ee
Moreover the $(\bf{3},\bf{2})^{-5/3}+(\bf{\bar{3}},\bf{2})^{+5/3}$ representation takes a factor $e^{2\pi i 5q/p}$
under the gauge transformation $U_\Gamma$. The condition (\ref{ModBoundCond}) then becomes
\ba\label{ModBoundCondLens}
   T_{k; m_1,m_2}(y) = e^{2\pi i(\ell+m r+5q)/p} \,T_{k; m_1,m_2}(y)
\ea
and the invariant eigenmodes are those satisfying the constraint $\ell+m r+5q=0 \,\,{\rm mod}\,\, p$.

If one wants the right normalization, in order to get an orthonormal base, the
$\sqrt{\frac{1}{2\pi^2}}$ factor has to be changed in the more general  
$\sqrt{\frac{1}{V}}$, where $V$ is the volume of $L(p,r)$. In what follows we will call $T$ 
the eigenmodes of Laplacian with appropriately modified normalization.

\subsection{Green's function}

Having the Laplacian eigenmodes on $L(p,r)$, we can compute the Green's function explicitly:
\be 
G(y_1,y_2)=\sum_{{\tiny  \begin{array}{c}  k;m_1,m_2\\constr\\ \end{array}}} \frac{1}{\lambda_k}\bar{T}_{k; m_1,m_2}(y_1) 
T_{k; m_1,m_2}(y_2)
\ee
where the sum over $\lambda_k=-k(k+2)$ and $\{ k,m_1,m_2  \}$ is constrained by 
$\ell+m r+5q=0 \,\,{\rm mod}\,\, p$ and $m_1$ and $m_2$ running from $-k/2$ to $k/2$ with integer step.
We implement these constraints by using the fact that
\be
\frac{1}{p}\sum_{w=1}^p e^{2\pi i w(5q+\ell+mr)/p}
\ee
is equals to one if and only if $\ell+m r+5q=0 \,\,{\rm mod}\,\, p$ and is zero otherwise.

So we can write:
\ba\label{GG5}
G(y_1,y_2)&=&\sum_{{\tiny  \begin{array}{c}  k\not=0;m_1,m_2\\unconstr\\ \end{array}}} \frac{1}{\lambda_k}\bar{T}_{k; m_1,m_2}(y_1)\left(
\frac{1}{p}\sum_{w=1}^p e^{2\pi i 5q w/p}  e^{2\pi i w(\ell+mr)/p}  T_{k; m_1,m_2}(y_2) \right) \nn\\
&=& \frac{1}{p}\sum_{w=1}^p  u^w \sum_{{\tiny  \begin{array}{c}  k\not=0;m_1,m_2\\unconstr\\ \end{array}}} \frac{1}{\lambda_k}
\bar{T}_{k; m_1,m_2}(y_1)  T_{k; m_1,m_2}(\Gamma^wy_2)\nn\\
&=& \frac{1}{p}\sum_{w=1}^p  u^w \frac{2\pi^2}{V}G_{\mathcal{S}^3}(y_1,\Gamma^wy_2)
\ea
where $u\equiv e^{2\pi i 5wq/p}$ and
\be
G_{\mathcal{S}^3}(y_1,y_2)\equiv \sum_{k\not=0;m_1,m_2} \frac{1}{\lambda_k}
\bar{T}_{k; m_1,m_2}^{\mathcal{S}^3}(y_1)  T_{k; m_1,m_2}^{\mathcal{S}^3}((y_2)
\ee
is the regulated Green's Function on the sphere ({\it e.i.} one neglects the zero mode in the sum and the modes have the appropriate normalization for
the sphere), which we will compute in a moment.

By using (\ref{DDgr}) and \cite{integrals} one gets:
\ba
G_{\mathcal{S}^3}(y_1,y_2) &=&  \frac{1}{2\pi^2}\sum_{k=1}^\infty\frac{k+1}{-k(k+2)} \frac{\sin[(k+1)d ]}{\sin[d ]} \nn\\
&=&-\frac{1}{2\pi^2}    \frac{1}{\sin[d]}\sum_{h=2}^\infty\frac{h}{h^2-1} \sin[h d ]       \nn \\
&=& -\frac{1}{4\pi\tan d}+\frac{1}{8\pi^2}+\frac{d}{4\pi^2 \tan d}
\ea
When we use it in order to compute (\ref{GG5}), we can neglect the constant $1/8\pi^2$ because it gives zero 
contribution: it factors out from the sum over $w$, which is so equal to zero since $5q \not = 0 {\rm mod}p$. So
\be\label{GG6}
G(y_1,y_2) = \sum_{w=1}^p u^w \frac{d(y_1,\Gamma^w y_2)-\pi}{4\pi^2 \tan d(y_1,\Gamma^w y_2)}
\ee

We note that if we use the formula (\ref{GG6}) for the Green's function on $L(2,1)=\mathcal{S}^3/\Bbb{Z}_2$, 
we actually get the same result as (\ref{GGRP3}).

\newpage

\section{Green's function on $\Bbb{H}^3$}

In order to compute the fundamental solution to the the Heat equation (\ref{HeatK}) 
on $\Bbb{H}^3$, we will use the formula given 
at page $150$ of \cite{Chavel}:
\be
H(y_1,y_2;t)= (4\pi t)^{-3/2}\,e^{-d^2(y_1,y_2)/4t} e^{-t}\frac{d(y_1,y_2)}{\sinh d(y_1,y_2)}  
\ee

We compute the following integral over $t$:
\ba
  \int_0^\infty \frac{dt}{t^{3/2}}e^{-d^2/4t} e^{-t} &=& 2^{3/2} d^{-1/2} K_{1/2}(d)\nn\\
  &=& 2^{3/2} d^{-1/2} \frac{e^{-d}(2\pi)^{1/2}}{2 d^{1/2}}\nn\\
  &=& 2 \pi^{1/2}\frac{e^{-d}}{d}
\ea
Where $K_\nu$ is the modified Bessel function.
So the Green's function is given by
\ba
  G_{\Bbb{H}^3}(y_1,y_2;t) &=& -\int_0^\infty dt \, H(y_1,y_2;t)\nn\\
  &=& -(4\pi)^{-3/2} \frac{d(y_1,y_2)}{\sinh d(y_1,y_2)} \,2 \pi^{1/2}\frac{e^{-d(y_1,y_2)}}{d(y_1,y_2)}\nn\\
  &=& -\frac{1}{4\pi}\frac{e^{-d(y_1,y_2)}}{\sinh d(y_1,y_2)}
\ea

\end{document}